\documentclass[nofootinbib,amsfonts,prd,aps,superscriptaddress]{revtex4-1}
\usepackage[colorlinks,allcolors=black]{hyperref}
\usepackage{subfig}
\usepackage{amsmath,amssymb,amsfonts,amsthm,mathrsfs}
\usepackage{cancel}
\usepackage{stackrel}
\usepackage{graphicx}
\usepackage{nameref}
\usepackage{filecontents}

\newcommand{\IFFC}{Instituto de F\'{\i}sica, Facultad de Ciencias, Igu\'a 4225, esq. Mataojo, 11400 Montevideo, Uruguay.}
\newcommand{\ULA}{Centro de F\'{\i}sica Fundamental, Facultad de Ciencias, ULA, Venezuela.}
\begin{document}

\title{Quantum fluctuating CGHS geometries and the information paradox}

\author{Rodrigo Eyheralde}
\affiliation{\IFFC}
\author{Rodolfo Gambini} 
\affiliation{\IFFC}
\author{Aureliano Skirzewski}
\email{askirz@gmail.com}
\affiliation{\IFFC}
\affiliation{\ULA}

\begin{abstract} 
We study Hawking radiation on the quantum spacetime generated by a quantum ingoing null shell in the 2d theory proposed by Callan-Giddings-Harvey-Strominger (CGHS). The quantum spacetime is a superposition of classical geometries with uncertainty in position and momentum  of the collapsing shell. The Hawking radiation spectrum is computed, revealing a non-thermal behaviour for finite time as well as a dependence on the shell's physical degrees of freedom. Hawking radiation's dependence on the collapsing shell becomes irrelevant in the late time approximation as we reach $i^+$ since the radiation's temperature depends exclusively on the cosmological constant. However, we argue that the information of the quantum state of a collapsing shell can be read from the Hawking radiation if we perform measurements at  $\mathcal{I_R^+}$ taking into account backreaction effects.
\end{abstract}

\maketitle

\section{Introduction}
The study of black hole evaporation is one of the most important problems of fundamental physics nowadays. It involves gravity, quantum field theory and thermodynamics in their full regimes. Hawking's calculation \cite{Hawking:1974sw} showing that black holes radiate with a thermal spectrum initiated the study of this phenomenon. However a complete understanding of the evaporation process is still lacking and important contradictions between some of the fundamental principles of physics, \cite{Hajicek:2002ny,Townsend:1997ku,Almheiri:2013hfa,Marolf:2017jkr,Unruh:2017uaw} see also p.49 \cite{Chakraborty:2017pmn} appear. The standard calculation of Hawking radiation assumes a fixed spacetime ignoring the fact that the black hole loses mass through the radiation and eventually evaporates \cite{Kuchar:1994zk}. Associated with the evaporation process appears the issue of information loss. The black hole evaporates in a thermal state characterized by only one number, its temperature, and the information about the matter that formed the black hole is apparently lost. 
Having a model calculation for the formation of a black hole and its evaporation including quantum effects induced by horizon fluctuations would be very useful to gain greater understanding of this process.

In a previous paper \cite{Eyheralde:2017jzd}, Hawking radiation on the quantum spacetime of a collapsing null shell in 3+1 dimensions was studied. The quantum
spacetime was constructed by superposing classical geometries associated with collapsing shells with uncertainty in their position and mass. Departures from thermality were observed in the radiation even though backreaction was not considered. The usual profile for the Hawking radiation as a function of frequency was recovered in the limit where the spacetime is classical. However, when quantum corrections were taken into account, it was shown that the profile of the Hawking radiation as a function of time contains at least part of the information about the initial state of the collapsing shell. 

Two-dimensional models of black holes were introduced some twenty five years ago \cite{Callan:1992rs,Giddings:1992ff,Strominger:1994tn,Schoutens:1993hu,Varadarajan:1995jj}. Even though they are simpler than the four-dimensional case, most of the previously mentioned problems are still present \cite{Ashtekar:2008jd,Ashtekar:2010qz,Modak:2016uwr,Guo:2018cgy,Ho:2018jkm}, and a complete quantum analysis of a collapsing system is still lacking. In this paper we start the study of a collapsing quantum shell of matter in two dimensions. New techniques for the computation 
of the Hawking radiation emitted by the quantum shell are developed. It is shown that in spite of the fact that the temperature of the classical two-dimensional black hole is universal and independent of the properties of the black hole, the radiation emitted by a quantum shell depends on its properties. This raises the expectation of being able to recover the information initially contained in the system.

\section{Collapsing Shell}
Here we study the collapse of a null shell. In order to do so, let us consider the action of CGHS's model coupled to a null shell and a massive scalar field 
\begin{equation}
S_{CGHS}+S_{Shell}+S_\eta=\frac{1}{2G}\int d^2x \sqrt{g}e^{-2\phi}\Big(R+4\partial_a\phi\partial^a\phi+4\lambda^2\Big)-m\int d\tau \sqrt{ -\dot q_a(\tau)\dot q^a(\tau)}-\int d^2x\sqrt{g}\frac{1}{2}\partial_a\eta\partial^a\eta
\end{equation}
At the zero mass limit, and considering the first order action written in terms of the momentum $p$  of the shell \cite{Louko:1997wc}, one gets 
\begin{equation}
S_{Shell}=\int dt \Big[p\dot q-\hat N\eta(p)p-\hat N^xp\Big], 
\end{equation}where $\eta(p)$ is the sign of $p$, and $ \hat N$ and $\hat N^x$ correspond to the lapse and shift functions of the metric, evaluated at the shell's position, $$g=-N^2dt^2+q(dx +N^xdt)^2.$$ Additionally, we have considered the coupling with a quantum scalar field $\eta$, whose quantum correlations allow us to take into account  the black hole radiation.

The variations of the action with respect to the metric lead to
\begin{equation}
0=\nabla_a\nabla_be^{-2\phi}-g_{ab}\nabla_c\nabla^ce^{-2\phi}-4e^{-2\phi}\partial_a\phi\partial_b\phi+2g_{ab}e^{-2\phi}\partial_c\phi\partial^c\phi+2g_{ab}\lambda^2 e^{-2\phi}+GT_{ab}
\end{equation}
where
\begin{equation}
T_{ab}=\frac{p_{(a}\dot q_{b)}}{\sqrt{g}}\delta(x^+-x_0^+)+\partial_a\eta\partial_b\eta-\frac{1}{2}g_{ab}\partial_c\eta\partial^c\eta .
\end{equation}
Using Kruskal coordinates (given by $g=-e^{2\rho}dx^{+}dx^{-}$) and assuming the shell comes from the point $x^+=x_0^+$ in $\mathcal{I_R^-}$, the only non zero components of the connection are $\Gamma^{\pm}{}_{\pm\pm}=2\partial_\pm\rho$ and the curvature is $R=8e^{-2\rho}\partial_{+}\partial_{-}\rho$. Thus, equations of motion for $g_{+-}$ and the dilaton $\phi$ are
\begin{eqnarray}\label{eomFree1}
-2\partial_{+}\partial_{-}\phi+4\partial_{+}\phi\partial_{-}\phi +\lambda^2e^{2\rho}&=&0\\\label{eomFree2}
-2\partial_{+}\partial_{-}\rho+4\partial_{+}\partial_{-}\phi -4\partial_{+}\phi\partial_{-}\phi -\lambda^2e^{2\rho}&=&0.
\end{eqnarray}
One can show by considering the addition of the previous equations and coordinate transformations that the system admits a gauge fixing  $\rho=\phi$ \cite{Callan:1992rs}.
The coupled CGHS/Shell's equations (for a collapsing shell with momentum $p<0$ and no classical modes for the scalar field $\eta$) become
\begin{equation}
0=\begin{pmatrix}
\partial_{-}^2e^{-2\rho}& -\partial_{+}\partial_{-}e^{-2\rho}-\lambda^2\\-\partial_{+}\partial_{-}e^{-2\rho}-\lambda^2 & \quad \partial_{+}^2e^{-2\rho}-2Gp\delta(x^{+}-x_0^{+})
\end{pmatrix}
\end{equation}
whose solution is given up to translations by
\begin{equation}\label{BH-shell}
e^{-2\rho}= \frac{GM_0}{\lambda}-\lambda^2x^{+}x^{-}+2Gp(x^{+}-x_0^{+})\Theta(x^{+}-x_0^{+}),
\end{equation}
where $M_0$ is an integration constant that will be associated to the mass of a previously existing black hole. In the following section we are going to identify the observable quantities that can be defined for this system for its quantization.

\subsection{Canonical formulation of CGHS's action}
In this section we introduce the canonical analysis of the dynamics of the CGHS model by following \cite{Kuchar:1994zk,Varadarajan:1995jj,Campiglia:2016fzp}. The CGHS's action \cite{Rastgoo:2013isa,Corichi:2016nkp} can be rewritten in a vielbein formulation
\begin{equation}
S_{CGHS}=\frac{1}{G}\int d^2x\Big\{-X_{I}\epsilon^{ab}(\partial_{[a}e_{b]}{}^{I}+\epsilon^{I}{}_{J}\omega_{[a}e_{b]}{}^{J})+\frac{1}{8}\Phi^2\partial_{[a}\omega_{b]}e\epsilon^{IJ}e^{a}{}_{I}e^{b}{}_{J}+\frac{1}{2}\eta^{IJ}ee_{I}{}^{a}e_{J}{}^{b}\partial_{a}\Phi\partial_{b}\Phi+\frac{1}{2}\lambda^2e\Phi^2\Big\}.\label{eq:act-tet1}
\end{equation}
with local Minkowski invariant tensors 
$\eta_{IJ}, \epsilon^{IJ}$, frames $e_a^I$, spin connection $\omega_a=e^\nu_J\epsilon^J{}_I(\partial_\mu e^I_\nu-\Gamma^\lambda_{\mu\nu}e^I_\lambda)$, dilaton $\Phi=2e^{-\phi}$ and Lagrange multipliers $X^I$.
By analogy with $3+1$ gravity as in \cite{Gambini:2008dy}, the spatial line element may be written as $ds^2=(E^\varphi)^2dx^2$ and $E^x = e^{-2\phi}$, a new set of variables can be chosen for the canonical formulation of the CGHS action \cite{Corichi:2016nkp}. 

At this point we follow	\cite{Corichi:2016nkp} and substitute the lapse and shift functions by $$N=\frac{(E^x)'}{E^\varphi E^x}\bar N,\text{ and } N^x=\bar N^x -\frac{K_\varphi}{E^\varphi E^x}\bar N$$ where $F':=\partial_x F=\partial_{+}F+\partial_{-}F$. We can then rewrite the constraints in the Hamiltonian formulation as
\begin{equation}\label{CGHS-COR}H(\bar N)=\frac{1}{G}\int dx\;\bar{N}
\left[\frac{1}{2}\frac{E^{x\prime2}}{E^{\varphi2}E^{x}}-2E^{x}\lambda^{2}-\frac{1}{2}\frac{K_{\varphi}^{2}}{E^{x}}- GpF(x^{+}_0)\Theta(x^{+}-x_0^{+})+2\lambda  G M_0\right]'
\end{equation}
for a collapsing shell with $p<0$, and 
\begin{equation}\label{CGHS-COR-Dif}D(N^x)=\frac{1}{G}\int dx\bar{N}^{x}\left(-U_{x}E^{x\prime}+E^{\varphi}K_{\varphi}'-p\delta(x^{+}-x_0^{+})\right),
\end{equation}
where the new set of canonical pairs $K_\varphi, E^\varphi,U_x,E^x,$ satisfy $\{K_\varphi(x), E^\varphi(y)\}=G\delta(x-y)$ and $\{U_x(x), E^x(y)\}=G\delta(x-y)$ and all the other brackets are trivial. They can be related to the  Kruskal coordinates representation of the black hole's metric (\ref{BH-shell}) setting  $K_{\varphi}=-\frac{1}{2}\frac{P_\Phi \Phi}{E^\varphi}$ (eq. (3.29) in \cite{Corichi:2016nkp}),  $x^{\pm}=t\pm x$, $\sqrt{q}=e^\rho$, $N=e^\rho$, $N^x=0$, and $\rho= \phi$. From our basic assumptions we can also determine the expressions for \begin{equation}\label{F}
P_\Phi=\frac{\sqrt{q}}{N}(N^x\partial_x\Phi-\partial_t\Phi)\quad \text{ and }\quad
F(x)=-\frac{K_\varphi}{E^\varphi E^x}+\frac{(E^x)'}{(E^\varphi)^2 E^x }.
\end{equation}

\subsection{CGHS/Shell observable quantities}\label{sec:quantum_shell}
We are now ready to identify the Dirac observables.
From the gravitational part of the Hamiltonian, let us define  \begin{equation}
H_g(x)=\frac{1}{2}\frac{E^{x\prime2}}{E^{\varphi2}E^{x}}-2E^{x}\lambda^{2}-\frac{1}{2}\frac{K_{\varphi}^{2}}{E^{x}}.
\end{equation} Integrating by parts in (\ref{CGHS-COR}) and taking into account the boundary terms at spatial infinity, we can redefine the total Hamiltonian as
\begin{equation}\label{CGHS-COR+Campi}H(\bar N)=-\frac{1}{G}\int dx\;\bar{N}'
\left[H_g+ G|p|F(x^{+}_0)\Theta(x^{+}-x_0^{+})+2\lambda  G M_0\right]+
\bar N^{+}2\lambda M^{+}+
\bar N^{-}2\lambda M^{-}
\end{equation} where $M^{-}=M_0$ is the mass of a previously existing BH and $M^{+}=M_0+\frac{|p|F(x^{+}_0)}{2\lambda}$ is the mass of the black hole after the collapse of the shell.
For simplicity, we will restrict our attention to the ${\cal I}^+_R$ observables. 
$M^{+}$ can be evaluated on the shell's BH solution (\ref{BH-shell}) $$M^{+}=M_0-\frac{pF(x^{+}_0)}{2\lambda}=M_0- p\lambda x_0^{+},$$ which tells us that the collapse of the shell increases the black hole's mass. 
If $M_0=0$ there is no black hole spacetime until $x^{+}\geq x^{+}_0$ and the mass turns to $M=-p\lambda x^{+}_0$ after the  collapse of the shell. The function $F(x)$ given in (\ref{F}) can be used to define two Dirac observables.
One of them is the mass of the shell \begin{equation}\label{M}
M=-\frac{pF(x_0^{+})}{2\lambda}.
\end{equation} To prove it, let's compute it's Poisson bracket with the total Hamiltonian. We start by computing  \begin{equation}
\Big\{F(x),H_g(y)\Big\}=\Big\{F(x),H_g(y)\Big\}_{grav}+\cancelto{0}{\Big\{F(x),H_g(y)\Big\}_{Shell}}=GF^2(x)\delta(x-y),
\end{equation} where $grav$ and $shell$ reffer to the restriction of the Poisson brackets to the gravitational and shell degrees of freedom respectively.
 It follows that \begin{equation}\label{DiracM}
\Big\{-\frac{pF(x_0^{+})}{2\lambda},H_g(x)-GpF(x^+_0)\Theta(x^+-x^+_0)\Big\}=\Big\{-\frac{pF(x_0^{+})}{2\lambda},H_g(x)\Big\}_{grav}
\!\!\!+\Big\{-\frac{pF(x_0^{+})}{2\lambda},-GpF(x^+_0)\Theta(x^+-x^+_0)\Big\}_{Shell}=0,
\end{equation} showing that $M$ is in fact a Dirac observable.

Additionally, we can recognize another Dirac observable \cite{Campiglia:2016fzp} of the coupled system \begin{equation}\label{V}
V=-\int_{x^+_0}^\infty dy\frac{2\lambda}{F(y)}
\end{equation} whose Poisson bracket with the total Hamiltonian (as in (\ref{DiracM})) can also be shown to be zero. The observable $V$ is associated with the Eddington--Filkenstein coordinate $v$ of an observer at $\cal{I}^-$ from which the shell is incoming or exiting. It can also be shown that $V$ is canonically conjugate to $M$ since \begin{equation}
\{M,V\}=\cancelto{0}{\{M,V\}_{grav}}+\{M,V\}_{Shell}=1.
\end{equation}

In what follows we will use these Dirac observables for the quantum description of the shell and the induced spacetime. Observable quantities can be recognized by their dependence on $M$, $V$ and classical parameters such as the coordinates at $\mathcal{I}^\pm_{R/L}$.   

\section{Hawking Radiation }

The Hawking radiation is a flux of particles released by black holes due to quantum effects near the event horizon. Here, it will be expressed in terms of the expectation value of the density number operator $\hat N_{out}$ of a massless quantum field. In the CGHS/Shell model coupled to a quantum massless scalar field, radiation is observed at $\mathcal{I}_R^{+}$ due to the effects of dilatations on the flux of particles of the scalar field operator $\eta(x,t)$, from  $\mathcal{I}_L^{-}$ to $\mathcal{I}_R^{+}$. In fact, since the free scalar field equation is 
$$\partial_+\partial_-\eta=0$$
the free scalar field operator is decomposed into right and left moving parts $\eta(x,t)=\eta_{R}(x^{-})+\eta_{L}(x^{+})$ that evolve independently. Hawking radiation is determined only by right moving initial data. Therefore, from this point on, we will pay attention only too $\eta_{R}(x^{-})$ and the observables defined from it. In the $3+1$ BH the geometric optics approximation is needed to do this separation but here it arises naturally. In $\mathcal{I}_{L}^{-}$ the field operator can be represented as
\begin{equation}\label{etaScrI-}
\eta_R^{in}(x,t)=\int_0^\infty \!\!\!\!\!d\omega \Big\{f_\omega(x_{in}^{-}) a_{\omega}+f^*_\omega(x_{in}^{-}) a^\dagger_{\omega}\Big\},
\end{equation}
where $f_\omega(x)$ are a basis of classical solutions of the wave equation, labeled by $\omega$ and normalized a la Dirac on $\mathcal{I}_L^{-}$ with the norm conserved by the Klein Gordon's equation
\begin{equation}\label{NormScrI-}
(f_\omega,f_{\omega'})=-i\int_{\mathcal{I}_L^{-}}\!\!\!\!\!dx_{in}^{-}\Big(f_\omega\partial_{-}f^*_{\omega'}-\partial_{-}f_\omega f^*_{\omega'}\Big).
\end{equation}
A natural plane wave basis is the one given  by $f_\omega(x_{in}^{-}(\sigma_{in}^{-}))=\frac{1}{\sqrt{4\pi\omega}}e^{-i\omega\sigma_{in}^{-}}$ for right moving modes, with $-\lambda x_{in}^{-}= e^{-\lambda\sigma_{in}^{-}}$ at $\mathcal{I}^-_L$.

Aside from the natural separation in right and left moving modes, another remarkable property of this model is that the density number of particles can be determined at any null surface with constant $x^+=x^+_m$ for a constant quantum state $|\Phi(x^+_m)\rangle=|0_{in}\rangle$. Unlike on 3+1 models, a natural decomposition of the scalar field in positive and negative frequency modes can be done in any constant $x^+=x_m^+$, where 
\begin{equation}\label{coords}
x^-=x_{in}^--\frac{2Gp}{\lambda^2}+\frac{2Gpx_0^+}{\lambda^2x_m^+}.
\end{equation}
In particular, in $\mathcal{I}_R^+$
\begin{equation}
x_{out}^-=x_{in}^--\frac{2Gp}{\lambda^2}.
\end{equation} 
We will start by studing Hawking radiation as seen in this assymtotic region and then extend the results inside the bulk.
\subsection{The expectation value of the number of particles}
We are now going to describe the evolution of the scalar field in order to establish the radiation that escapes from the black hole and reaches $\mathcal{I}^+_R$. As we mentioned before, it is possible to find a set of normalized functions as in (\ref{NormScrI-}) at $\mathcal{I}_R^{+}$ and write the scalar field as in (\ref{etaScrI-})
\begin{equation}\label{etaScrI+}
\eta_R^{out}(x,t)=\int_0^\infty \!\!\!\!\!d\omega \Big\{f_\omega( x_{out}^{-}) b_{\omega}+f^*_\omega(x_{out}^{-}) b^\dagger_{\omega}\Big\}.
\end{equation}
Using this, we can define the expectation value of the number of particles, either on $\mathcal{I}^-_L$ or on $\mathcal{I}^+_R$, for a frequency $\omega$. 
While the first one can be computed by taking the mean value of 
$$\hat N_{in}({\omega})=\hat a_{\omega}^\dagger\hat a_{\omega}$$ 
(and is obviously zero for the 'in' vacuum $|0_{in}\rangle$ given by $\hat a_{\omega}|0_{in}\rangle=0$ for all $\omega$), the second is obteined as the expectation value of $$\hat N_{out}({\omega})=\hat b_{\omega}^\dagger\hat b_{\omega}.$$ 
 The known result is that it does not vanish on the quantum state $|0_{in}\rangle$ because this not a zero particles state at $\mathcal{I}^+_R$. The result is the observation of Hawking radiation at $\mathcal{I}^{+}_R$. To comute it we use the identification
$$\eta_R^{out}( x_{out}^-)=\eta_R^{in}(x_{in}^-(x_{out}^-))$$
for the right moving modes in both asymtotically flat regions. The annihilation operator of the $\eta$ fundamental excitations in $\mathcal{I}^+_R$ is
\begin{equation}
\hat b_{\omega} :=(f_\omega(x_{out}^{-}), \eta_{out})_{\mathcal{I}_R^+}=(f_\omega(x_{out}^{-}(x_{in}^-)), \eta_{in}(x_{in}^-))_{\mathcal{I}_L^-}=\int_0^\infty d\nu\Big\{\alpha^*_{\omega\nu}\hat a_{\nu}-\beta^*_{\omega\nu}\hat a^\dagger_{\nu}\Big\},
\end{equation} 

with Bogoliubov coefficients
\begin{equation}
\alpha_{\omega\nu}=\Big(f_\omega( x_{out}^{-}(x_{in}^-)),f_{\nu}(x_{in}^{-})\Big);\ \beta_{\omega\nu}=-\Big(f_\omega(x_{out}^{-}(x_{in}^-)),f^*_{\nu}(x_{in}^{-})\Big)
\end{equation}
After its transit through near BH geodesics, the right moving modes get scaled and the $\beta_{\omega\nu}$ are not vanishing. Because of that, the annihilation operator $\hat b_{\omega}$ act on the vacuum state with non trivial effects. To compute this Bogoliubov coefficient we set $x_{out}^{-}(x_{in}^{-}) = x_{in}^{-}-\frac{2Gp}{\lambda^2}$ and $-\lambda x_{out}^{-}=e^{-\lambda\sigma_{out}^{-}}$ so, 
\begin{equation}\label{Bogoliubov}
\beta_{\omega\nu}=-\frac{1}{2\pi\lambda}\Big(\frac{-2Gp}{\lambda}\Big)^{i\frac{\omega+\nu}{\lambda}}\sqrt{\frac{\omega}{\nu}}B(-i\frac{\omega+\nu}{\lambda},i\frac{\omega}{\lambda})
\end{equation}
where $B(a,b)=\frac{\Gamma(a)\Gamma(b)}{\Gamma(a+b)}$, is the beta function. Now, instead of computing the number density opperator, is useful to introduce the so called density matrix operator $\hat b^\dagger_{\omega_1}\hat b_{\omega_2}$ which has the number density operator as diagonal but also has off-diagonal terms. Its expectation value is \begin{equation}\label{rho12}
\rho(\omega_1,\omega_2)=\langle 0_{in}|\hat b^\dagger_{\omega_1}\hat b_{\omega_2}|0_{in}\rangle =\frac{\hbar\sqrt{\omega_1\omega_2}}{(2\pi\lambda)^2}\Big(-\frac{2Gp}{\lambda}\Big)^{i\frac{\omega_1-\omega_2}{\lambda}}\frac{\Gamma(i\frac{\omega_1}{\lambda})\Gamma(-i\frac{\omega_2}{\lambda})}{\pi} K (\omega_1,\omega_2)
\end{equation} where
\begin{equation}
K (\omega_1,\omega_2)= \int_0^\infty \frac{d\nu}{\lambda} \sinh(\pi \frac{\nu}{\lambda}) \Gamma(-i(\frac{\omega_1 +\nu}{\lambda}))\Gamma(i(\frac{\omega_2+\nu}{\lambda})).\label{K}
\end{equation}
In the above expression for $ K(\omega_1,\omega_2)$ we have left the explicit integral as we have not found an analytic expression for it. However, it should be pointed out that (\ref{K}) becomes divergent for $\omega_1=\omega_2$, 
it is this divergence what makes thermal behavior to dominate over other contributions to the number of particles.

The integral we have to perform involves the Gamma function.
On the imaginary axis the Gamma function can be expressed as \cite{GammaLowfreq:Online,GammaHighFreq:Online} $$\Gamma(iy)=\sqrt{\frac{\pi}{y\sinh (\pi y)}}e^{i\varphi(y)}$$ with $$\varphi(y)=y\ln(|y|) -\frac{ \pi }{4}sgn(y) -y+y\sum_{m=1}^\infty\frac{(-1)^mB_{2m}}{2m(2m-1)y^{2m}}=-\frac{\pi}{2}sgn(y)-\gamma y-y\sum_{m=1}^\infty\frac{(-1)^{m}\zeta(2m+1)y^{2m}}{2m+1},$$  where $B_{2m}$ are Bernoulli numbers $B_{2m}=\{\frac{1}{6},-\frac{1}{30},\frac{1}{42},-\frac{1}{30},\dots\}$ and $\zeta(2m+1)$ is the Riemann zeta function.
For small and big values of $y$, a valid approximation can be obtained from a truncation of the power series expansion proposed above.
The density matrix can now be rewritten as
\begin{equation}
\rho(\omega_1,\omega_2)=\frac{\hbar}{2\pi\lambda}\int_0^\infty\!\!\! {d\nu}
\sqrt{\frac{1}{\omega_1+ \nu}\Big(\frac{1}{e^{2\pi\frac{\omega_1}{\lambda}}-1}-\frac{1}{e^{2\pi\frac{\omega_1+\nu}{\lambda}}-1}\Big)}
\sqrt{\frac{1}{\omega_2+ \nu}\Big(\frac{1}{e^{2\pi\frac{\omega_2}{\lambda}}-1}-\frac{1}{e^{2\pi\frac{\omega_2+\nu}{\lambda}}-1}\Big)}
e^{iF(\omega_1,\omega_2, \nu,\sigma_0)},
\end{equation} 
where 
$$ F(\omega_1,\omega_2, \nu,\sigma_0)=\varphi(\frac{\omega_1}{\lambda})+\varphi(\frac{\omega_2+ \nu}{\lambda})-\varphi(\frac{\omega_1+ \nu}{\lambda})
-\varphi(\frac{\omega_2}{\lambda})
-(\omega_1-\omega_2)\sigma_0.$$ It is clear in this representation that the particles density $N(\omega)=\rho(\omega,\omega)$ is infinite because $F(\omega,\omega,\nu,\sigma_0)=0$ and the integral contains a logarithmic divergence. The reason for this result is the use of an improper basis with perfectly defined frequencies to calculate the density matrix. This way we are including the contributions to the radiation from all times. Additionally, since we do not consider back-reaction, the total emission of energy also adds up to infinity.

In order to define time dependent density of particles $N_\omega(\sigma_{out}^{-})$ we will use two different approaches. First we will use the Wigner's Functional \cite{mecklenbräuker1997wigner} as defined by
\begin{equation}
N_\omega(\sigma_{out}^{-})=\int d\omega'\rho{(\omega+\frac{1}{2}\omega',\omega-\frac{1}{2}\omega')} e^{i\omega'\sigma_{out}^{-}}\text{ with }\ \ \rho(\omega_1,\omega_2)=\frac{1}{2\pi}\int d\sigma_{out}^-  N_{\frac{\omega_1+\omega_2}{2}}(\sigma_{out}^{-})e^{-i\sigma_{out}^-(\omega_1-\omega_2)}
\end{equation}
where $\omega_1=\omega+\frac{1}{2}\omega'$ and $\omega_2=\omega-\frac{1}{2}\omega'$. Then we will consider wave packets in frequency and time as it was done originally by Hawking and compare the two.

\subsection{Density matrix's approximations}

Using the saddle points approximation \cite{wong2001asymptotic} of the Wigner distribution \cite{mecklenbräuker1997wigner} in the high frequency limit we can estimate the value of the integral as 
\begin{equation}
N_\omega(\sigma_{out}^-)=\int_0^\infty d\nu\int_{-2\omega}^{2\omega}d\omega'g(\omega,\nu,\omega')e^{if(\omega,\nu,\omega',\sigma_{out}^--\sigma_0)}=\frac{2\pi}{\sqrt{\det{A}}}g(\omega,\nu_0,\omega'_0)e^{if(\omega,\nu_0,\omega'_0,\sigma_{out}^--\sigma_0)},
\end{equation}
where $\sigma_0=-\frac{1}{\lambda}\ln(-\frac{2Gp}{\lambda})$ and $A$ is the Hessian of $f$ for the variables $\nu,\omega'$ at the stationary (critical) point of $f$.

The Wigner's Functional for the density matrix is then
\begin{equation}
N_\omega(\sigma_{out}^{-})=\int d\omega' \rho{(\omega+\frac{1}{2}\omega',\omega-\frac{1}{2}\omega')} e^{i\omega'\sigma_{out}^{-}}.
\end{equation}
Using $f(\omega,\nu,\omega',\sigma_{out}^--\sigma_0)=F(\omega+\frac{\omega'}{2},\omega-\frac{\omega'}{2}, \nu,\sigma_0-\sigma_{out}^-)$ we can find the critical point $(\nu_0,\omega'_0)$ for the resolution of the integral using  the stationary phase approximation with $\partial_\nu f(\omega,\nu_0,\omega'_0,\sigma_{out}^--\sigma_0)=0$ and $\partial_{\omega'} f(\omega,\nu_0,\omega'_0,\sigma_{out}^--\sigma_0)=0$. For this we get 
\begin{equation}
\begin{pmatrix}
\lambda\partial_\nu f
\\ 
2\lambda\partial_{\omega'} f
\end{pmatrix}=\begin{pmatrix}
\varphi'(\frac{\omega-\frac{\omega'_0}{2}+ \nu_0}{\lambda})
-\varphi'(\frac{\omega+\frac{\omega'_0}{2}+ \nu_0}{\lambda})
\\
\varphi'(\frac{\omega+\frac{\omega'_0}{2}}{\lambda})
-\varphi'(\frac{\omega-\frac{\omega'_0}{2}+ \nu_0}{\lambda})
-\varphi'(\frac{\omega+\frac{\omega'_0}{2}+ \nu_0}{\lambda})
+\varphi'(\frac{\omega-\frac{\omega'_0}{2}}{\lambda})
+2\lambda(\sigma_{out}^--\sigma_0).
\end{pmatrix}
\end{equation}
Since $\varphi'(x+y)-\varphi'(x-y)$ is an odd function in $y$ we conclude that $\omega'_0=0$, thus we are left with only one equation $$\lambda(\sigma_{out}^--\sigma_0)=\varphi'(\frac{\omega+\nu_0}{\lambda})-\varphi'(\frac{\omega}{\lambda}).$$ The equation above determines  $\nu_0$, the critical value of the frequency of the modes at $\mathcal{I}_R^-$ that contribute most to the density matrix. However, the above equation can be solved exclusively for $ \sigma_{out}^--\sigma_0\geq 0$ as the minimum of $\varphi'(\frac{\omega+\nu_0}{\lambda})-\varphi'(\frac{\omega}{\lambda})$ can be found at $\nu_0=0$ for all values of $\omega$. 
Thus
\begin{eqnarray}
N_\omega(\sigma_{out}^{-}\!)\!\!&=&\int_{\mathbb{R}}d\omega' \rho{(\omega+\frac{1}{2}\omega',\omega-\frac{1}{2}\omega')} e^{i\omega'\sigma_{out}^{-}}\\
&=&\hbar\Bigg\{
\frac{1}{e^{2\pi\frac{\omega}{\lambda}}-1}
-\frac{1}{e^{2\pi\frac{\omega}{\lambda}e^{\lambda(\sigma_{out}^--\sigma_0)}}-1}\Bigg\}\Theta(\sigma_{out}^{-}-\sigma_0),\label{rho-psi}
\end{eqnarray}where the stationary point has been found in $\omega'_0=0$ and  $\nu_0+\omega\simeq\omega e^{\lambda( \sigma_{out}^--\sigma_0)}+O(\omega^{-1})$. 
Here the first term corresponds to thermal radiation and the second provides deviations from thermality but its effects last for short time after the radiation starts.
We can't get a closed expression for $\nu_0$  but further corrections in the high frequency limit $\frac{\lambda}{\omega}$ give $\nu_0+\omega\simeq(\omega -\frac{\lambda^2}{2\omega}) e^{\lambda( \sigma_{out}^--\sigma_0)}+O((\frac{\lambda}{\omega})^{3})$.
Additionally, the solution to the integral could be further corrected by incorporating more terms to the stationary point approximation of the integral. Ideally, the energy radiated would coincide with the one obtained by considering the conformal anomaly (\ref{EnergyDensityScalar}), which generates a non trivial transformation of the energy flux from being exactly zero at $\mathcal{I}_L^-$ to a non-zero expression at $\mathcal{I}_R^+$.

\subsection{Fast Fourier Transform}
We also perform an alternative set of approximations following the analysis performed by \cite{Eyheralde:2017jzd}.
If  ${\sigma}_n^{-}=\frac{2\pi n}{\epsilon}$ and $\omega_j=(j+\frac{1}{2})\epsilon$, then the fast Fourier transform \cite{cohen1995time} can be expressed as
\begin{equation}
N^{FFT}_{\omega_j}{({\sigma}^{-}_n)}:=\frac{1}{\epsilon}\int_{j\epsilon}^{(j+1)\epsilon}d\omega_1\int_{j\epsilon}^{(j+1)\epsilon}d\omega_2 e^{i{\sigma}^{-}_n(\omega_1-\omega_2)} \rho(\omega_1,\omega_2),
\end{equation}
it represents the radiation in a window of time $\sigma^- \in(\frac{2\pi n}{\epsilon},\frac{2\pi (n+1)}{\epsilon})$.
 We rewrite this expression in terms of ${\omega}$ and $\omega'$, and the leading terms give 
\begin{equation}
N^{FFT}_{\omega_j}{({\sigma}^{-}_n)}= \frac{1}{\epsilon}\int_{-\epsilon}^{\epsilon}d\omega' e^{i{\sigma}^{-}_n\omega'}\int_{j\epsilon+\frac{|\omega'|}{2}}^{(j+1)\epsilon-\frac{|\omega'|}{2}}d\omega\ \ \rho_0(\omega+\frac{1}{2}\omega',\omega-\frac{1}{2}\omega').
\end{equation}
The diagonal part of the density matrix can be recognized to contribute the most to the particle density, however an approximation is imperative to be able to solve the integrals. Assuming $\arg(\Gamma(iy))\simeq y\ln(y/e)-\frac{\pi}{4}$ (which is good for high frequencies) and setting $(1-e^{-2\pi\frac{\nu}{\lambda}})/(1-e^{-2\pi\frac{\nu+\omega_j}{\lambda}}) \simeq 1$ inside the integral in $\nu$\footnote{Since $\omega=\frac{1}{2}(\omega_1+\omega_2)$ is the mean frequency of the incoming wave, this is equivalent to considering the
late-time limit} we get
\begin{equation}
N^{TF}_{\omega_j}{({\sigma}^{-}_n)}\simeq \frac{\hbar}{2\pi}
\frac{1}{e^{2\pi\frac{(j+\frac{1}{2})\epsilon}{\lambda}}-1}
\int_{-\epsilon}^{\epsilon}d\omega' e^{i({\sigma}^{-}_n-\sigma_0)\omega'}(1-\frac{|\omega'|}{\epsilon}) \Big(\pi\delta(\omega')+p.v.\Big(\frac{1}{i\omega'}\Big)\Big).
\end{equation}
This replicates for the CGHS case the results obtained by Eyheralde et al. \cite{Eyheralde:2017jzd}, resulting in
\begin{equation}\label{rhoTF}
N^{FFT}_{\omega_j}{({\sigma}^{-}_n)}\simeq \hbar
\frac{1}{e^{2\pi\frac{\omega_j}{\lambda}}-1}\Bigg(\frac{1}{2} +\frac{1}{\pi}Si(\epsilon(\sigma_n^--\sigma_0))+\frac{1}{\pi}\frac{\cos(\epsilon(\sigma_n^--\sigma_0))-1}{\epsilon(\sigma_n^--\sigma_0)}\Bigg),
\end{equation}
which is basically  thermal radiation, with temperature  $T\propto \lambda$, times a step function. This corresponds to the solution (\ref{rho-psi}) in the range $\lambda({\sigma}_{out}^--\sigma_0)>>1$.
In fact, with the Time-Frequency analysis we are able to predict the right starting time for the radiation $\sigma_{out}^-=\sigma_0$ but the Wigner functional for the density matrix provides further corrections to the spectrum, i.e. a bigger delay is observed in the amplitude of the radiation with smaller frequencies. This delay is accounted in first approximation by the substraction of a part of the spectrum that could be described as a thermal radiation too, with decreasing temperature $\tilde T\propto\lambda e^{-\lambda(\sigma_{out}^--\sigma_0)}$.

\section{The infalling quantum shell and Hawking radiation}
To take into account the quantum nature of the shell and the spacetime determined by it, we follow the reduced phase space quantization procedure introduced in section \ref{sec:quantum_shell}. For this reason, we promote the observable quantities $V$ and $M$, defined in (\ref{V}) and (\ref{M}), to operators satisfying the Heissenberg algebra
$$[\hat M,\hat V]=i\hbar.$$
As we have shown, the density matrix operator depends exclusively on $\sigma_0=-\frac{1}{\lambda}\ln\Big(-\frac{2Gp}{\lambda}\Big) $, which in turn depends on the quantum observables through $\hat p=-\widehat{Me^{-\lambda V}}=-e^{-\frac{\lambda\hat V}{2}} \hat  M e^{-\frac{\lambda\hat V}{2}}$.
$\sigma_0$ marks the initial time for the detection of Hawking radiation at $\mathcal{I}_R^+$ according to expression (\ref{rho-psi}). It is important at this point to note that, in principle, the choice of $\sigma_{out}^-$ depends on the properties of the collapsing null shell, i.e. it depends of $V$ and $M$. However, in this quantization we identify the conformal null infinity $\mathcal{I}_R^+$ of black hole spacetimes with different values of $\hat p$ by taking the coordinate $\sigma_{out}^-$ as a common coordinate for $\mathcal{I}_R^+$ independently of the properties of the shell. This way $N_\omega(\sigma_{out}^{-})$ depends on $\sigma_{out}^-$ as a parameter.

\subsection{Energy density }

The quantization of a scalar field in a curved background \cite{Parker:2009uva,Schwartz:2013pla,Birrell:1982ix} offers many difficulties but some of them can be overcome when the field is coupled to the CGHS model because in that case the energy momentum tensor is explicitly renormalizable.  In particular, the symmetries of the coupled system and its quantum anomaly provide means to compute the density flux of energy that reaches null infinity due to the quantum fluctuations of the scalar field. 
By taking into account the quantum anomaly of the conformal symmetry, the renormalized energy-momentum tensor can be written as \begin{equation}\label{ConformalEnergyDensityFlux}
T_{ab}=-p\delta(x^+-x_0^+)\begin{pmatrix}
1&0\\0& 0
\end{pmatrix}+\begin{pmatrix}
\langle:(\partial_+\eta)^2:\rangle &0\\0& \langle:(\partial_-\eta)^2:\rangle
\end{pmatrix}
-\frac{\hbar}{12\pi}\begin{pmatrix}
\partial_+\rho\partial_+\rho-\partial^2_+\rho&\partial_+\partial_-\rho\\\partial_+\partial_-\rho& \partial_-\rho\partial_-\rho-\partial^2_-\rho\end{pmatrix}.
\end{equation}
Traditionally, the energy density of the scalar field that escapes to $\mathcal{I}^+_R$ is obtained by considering a conformal transformation that makes $\langle:(\partial_\pm\eta)^2:\rangle$ different from zero. Using this conformal anomaly \cite{Callan:1992rs,Fabbri:2005mw} the energy that reaches  $\mathcal{I}^+_R$ at time $\sigma_{out}^-$ is 
\begin{equation}\label{EnergyDensityScalar}
T_{--}(\sigma_{out}^-)=\frac{\hbar\lambda^2}{48\pi}\Bigg(1-\frac{1}{(1+\frac{P(x^+_f)}{\lambda}e^{\lambda\sigma_{out}^-})^2}\Bigg).
\end{equation}
Here we follow an alternative procedure. From expression (\ref{rho-psi}) we compute the energy density at $\mathcal{I}_R^+$ as
\begin{equation}\label{EnergyDensityShell}
T_{--}^p(\sigma_{out}^-)=\int_0^{+\infty}d\omega \hbar\omega N_\omega(\sigma_{out}^{-})=\frac{\hbar\lambda^2}{48\pi}\Bigg(1-\frac{1}{\Big(\frac{-2Gp}{\lambda} e^{\lambda\sigma_{out}^-}\Big)^2}\Bigg)\Theta(\frac{-2Gp}{\lambda}-e^{-\lambda\sigma_{out}^-}).
\end{equation}
Obviously there are significant differences between the two expressions. In the first one, radiation is coming all the way from $-\infty$ and there is a sharp beginning in the second one. However, we expect this difference around $\sigma^-_{out}=-\frac{1}{\lambda}\ln(-\frac{2Gp}{\lambda}), $ to be smoothed by corrections to the saddle point approximation considered in the derivation of (\ref{EnergyDensityShell}). Although approximate, our result nonetheless gives us insight into the region of the spacetime where most of the  Hawking radiation is being produced.
\begin{figure}[htp] \centering{
\includegraphics[scale=0.8]{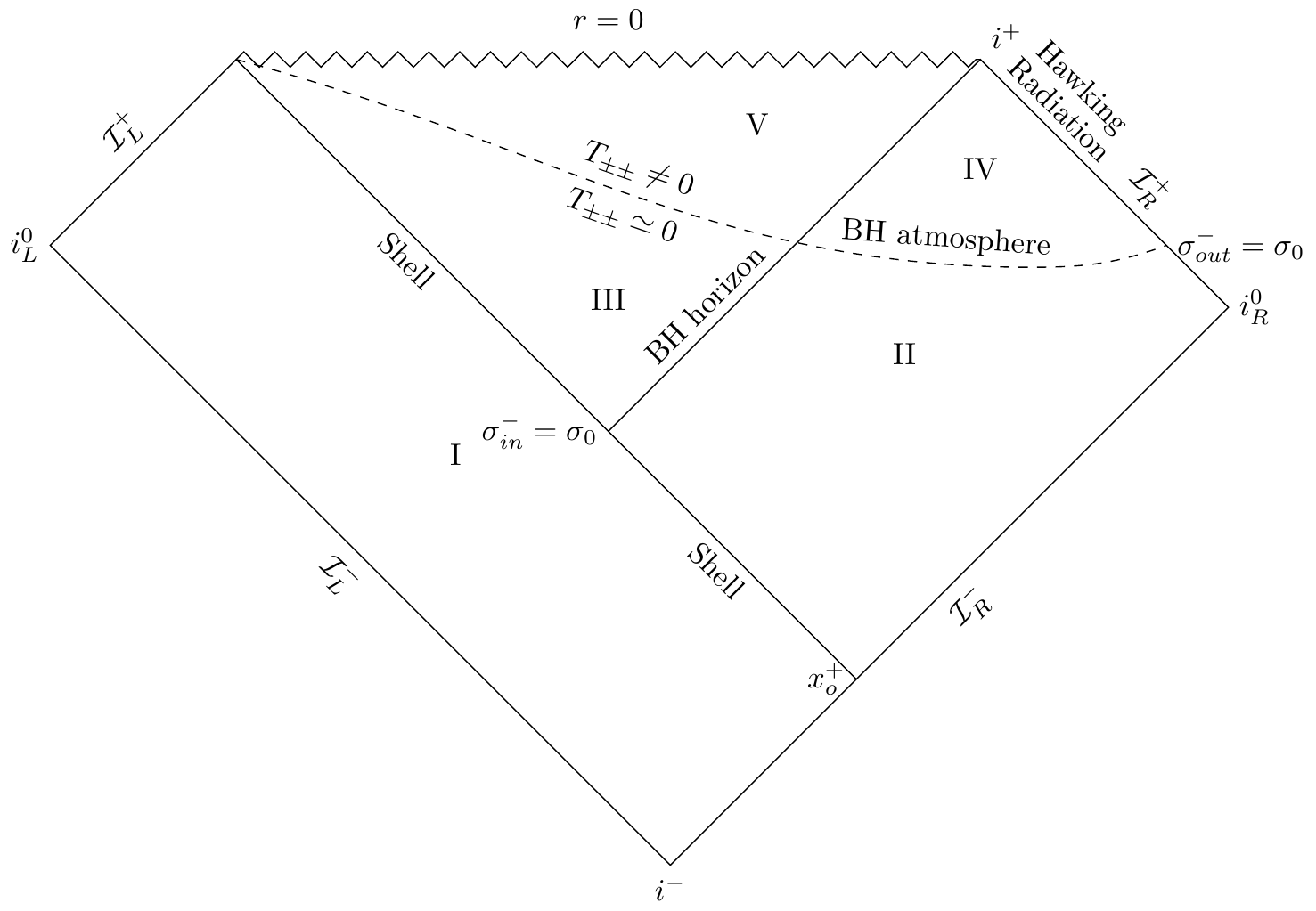}}
\caption{Region I represents a flat region of the spacetime, in II gravitational effects are important but BH radiation is not intense, in the region IV is where most of the Hawking radiation is produced and regions III and V represent the interior of the black hole, but differ by the mean energy momentum flux. Equation (\ref{EnergyDensityShell}) marks the end point of the dashed line at $\mathcal{I}_R^+$,  where  the energy density of BH radiation suddenly increases. The rest of the dashed line can be justified by equation (\ref{ParticleFlux})}
\label{fig:PwoBR}\end{figure} 
\subsection{Eigenstates of the Energy density operator}
As we have seen, the energy radiated at certain $d\sigma_{out}^-$ is a function of $-p$, whose quantization has to be performed in terms of the  Dirac observables $\hat M$ and $\hat V$ that we have previously studied. The whole quantization can make sense only if we restrict to the negative eigenvalues of $p=-\widehat{Me^{-\lambda V}}=-e^{\frac{-\lambda \hat V}{2}} \hat M e^{-\frac{\lambda \hat V}{2}}$, but this doesn't lead to a self-adjoint operator because the spectral theorem doesn't apply. 
Instead, we study the spectrum of $\hat p^2=(e^{-\frac{\lambda \hat V}{2}}\hat M e^{-\frac{\lambda\hat V}{2}})^2$, a self-adjoint operator in the space of eigenvectors of the operator $\hat V$, where $\hat M=i\hbar\partial_v$. $\hat p^2$ has a continuum spectrum, and its eigenvectors are
\begin{equation}\label{eigenvectorP}
R_p(v)=\sqrt{\frac{2}{\pi\hbar}}e^{\frac{\lambda v}{2}}\cos(\frac{-p}{\lambda\hbar}e^{\lambda v}).
\end{equation}If the  initial state of the shell is given by an eigenstate of $\hat p^2$ the radiated energy would be given by  (\ref{EnergyDensityShell}), but it would neither be sharp in $V$ nor $M$. 
We may, for instance, study how would Hawking radiation look like for a specific state. An interesting case to study would be a state as sharp as possible in both $V$ and $M$ around the values $\bar v$ and $\bar M$, for instance, a Gaussian $$\langle v|\chi_{gaussian}\rangle=\sqrt{\frac{ \pi}{a}}e^{-\frac{a}{2}(v-(\bar v+\frac{i}{a\hbar}\bar M))^2}.$$
With such state we could expect Hawking radiation to be better represented by its mean value with respect to the Gaussian state. Of course, the Gaussian state would smooth out the abrupt start of the radiation, which would be different from zero at all points in $\mathcal{I}^+_R$.

Other types of wave functions could be proposed and we could study the different possible mean density energy fluxes, however it seems these wavefunctions  would lead to a statistical mixture of the density energy flux (\ref{EnergyDensityShell}) of a classical shell. No new information could be extracted from the collapsing shell.

An interesting choice of the shell's wavefunction is one in which the mean  energy density of the quantum shell (\ref{EnergyDensityShell}) corresponds to that one obtained through the conformal anomaly (\ref{EnergyDensityScalar}). If we call it $\chi(p)=\langle p|\chi\rangle$, the equation 
\begin{equation}
T_{--}(\sigma_{out}^-)=\langle\chi|T_{--}^p(\sigma_{out}^-)|\chi\rangle=\int dp T_{--}^p(\sigma_{out}^-)|\chi(p)|^2
\end{equation} can be solved taking into consideration that the equation looks as a convolution. We obtain \begin{equation}
|\chi(p)|^2=3\lambda\frac{\frac{P(x^+_0)}{-2Gp}}{(1+\frac{P(x^+_0)}{-2Gp})^4},
\end{equation} however, it can be checked that  this wave function is not normalizable and with some suitable choice of a phase they could form an orthonormal improper basis. 

\subsection{Measurements \& Information}\label{measurements}
One may wonder if it is possible to recover the whole information of the shell through repeated measurement of the quantum system in $\mathcal{I}^+_R$. 
We could measure the number of particles within some domain in ${\mathcal{I}}_R^+$ with frequencies in some range. 
The probability to get an eigenvalue would be given by the projector into a specific subspace. 
Since the energy flux is an always increasing function, its determination in $(\sigma_{out}^-,\sigma_{out}^-+d\sigma_{out}^-)$ collapses the state of the system to a given eigenstate of the energy density (\ref{EnergyDensityShell}), i.e. the eigenvectors of $\hat p^2$ (\ref{eigenvectorP}). 
The determination of the particle flux depends on the shell exclusively through the operator $\hat p^2$, and complementary information about the state of the shell escapes the grasp of the observers at ${\mathcal{I}}_R^+$. 

In order to recover complementary information about the state of the Shell we have computed the density matrix in the bulk, at a finite value of $x^+=\frac{1}{\lambda}e^{\lambda \sigma^+}=x^+_m=\frac{1}{\lambda}e^{\lambda \sigma_m^+} $. 
Here, $x^\pm_m$ is defined as the position where the measurement of the density matrix takes place. 
On this region, with null coordinates $x^+=x^+_{in}$ and (\ref{coords})
\begin{equation}
x^-=x^-_{in}-\frac{2Gp}{\lambda^2}+\frac{2Gpx^+_0}{\lambda^2 x^+_m}
\end{equation} 
the metric looks like
\begin{equation}
g=-\frac{dx^+dx^-}{-\lambda^2x^+x^-+\frac{2Gpx^+_0}{x^+_m}(x^+-x_m^+)}=-\frac{d\sigma^+ d\sigma^-}{1-2Gpx_0^+e^{\lambda\sigma^-}(e^{-\lambda\sigma^+_m}-e^{-\lambda\sigma^+})},
\end{equation} which is a freely falling frame in coordinates $\sigma^\pm$ around $\sigma^+\simeq \sigma_m^+$.
These coordinates enable us to define field excitations of $\hat \eta(x)$ with frequency $\omega$ on the null trajectory with $x_{in}^+=x^+=x_m^+$. 
This can be done explicitly in $1+1$ dimensions since the free  massless field equation and consequently the preserved inner product, are independent of the metric \cite{Birrell:1982ix}. 
Thanks to these properties we can perform a Fourier transform and have an interpretation of the field modes into creation and annihilation operators. 
Thus, the density matrix can be determined in this region by computing the Bogoliubov coefficients 
\begin{equation}
\beta^m_{\omega\omega'}=-\frac{1}{2\pi\lambda}\Big(-\frac{2Gp}{\lambda}+\frac{2Gpx^+_0}{\lambda x^+_m}\Big)^{i\frac{\omega+\omega'}{\lambda}}\sqrt{\frac{\omega}{\omega'}}B(-i\frac{\omega+\omega'}{\lambda},i\frac{\omega}{\lambda}).
\end{equation}
Unlike (\ref{Bogoliubov}), here appears a dependence on the mass of the shell $M=-\lambda x_0^+ p$. Since the calculations are virtually the same as in $\mathcal{I}^+_R$, we just write the result for the Wigner transform
\begin{equation}\label{ParticleFlux}
N^{m}_{\omega}{({\sigma}^{-})}\simeq \hbar
\Bigg(\frac{1}{e^{2\pi\frac{\omega}{\lambda}}-1}-\frac{1}{e^{2\pi\frac{\omega}{\lambda} e^{\lambda(\sigma^--\sigma^-_{atm})}}-1}\Bigg)\Theta(\sigma^--\sigma^-_{atm}),
\end{equation} where $\sigma_{atm}^-=-\frac{1}{2\lambda}\log(-\frac{2Gp}{\lambda}-\frac{2GM}{\lambda^2 x^+_m})^2$ marks the start of the BH radiation at $x^+=x_m^+$. This corresponds to
\begin{equation}x_{in}^-(atm)=\frac{4G}{\lambda^2}(p+\frac{M}{\lambda x^+_m})\end{equation}
and defines a region we will call {\it Atmosphere of the BH}. Consequently the radiation in $(x^+,x_{in}^-)$ coordinates could be expressed by
\begin{equation}\label{atmosphere}
T_{--}(x^+,x_{in}^-)\simeq\frac{\hbar}{48\pi}\Big(\frac{1}{\big(x_{in}^--\frac{2G}{\lambda^2}(p+\frac{M}{\lambda x^+})\big)^2}-\frac{1}{\big(\frac{2G}{\lambda^2}(p+\frac{M}{\lambda x^+})\big)^2}\Big)\Theta\Big(x_{in}^--\frac{4G}{\lambda^2}\big(p+\frac{M}{\lambda x^+}\big)\Big).
\end{equation}
From the conformal anomaly (\ref{ConformalEnergyDensityFlux}) the analogous expression is
\begin{equation}\label{WithoutAtmosphere}
T_{--}(x^+,x_{in}^-)=\frac{\hbar}{48\pi}\Big(\frac{1}{\big(x_{in}^--\frac{2G}{\lambda^2}(p+\frac{M}{\lambda x^+})\big)^2}-\frac{1}{(x_{in}^-)^2}\Big),
\end{equation}
and these expressions converge to each other close to the singularity and very far from it.
The dependence of the radiation on the operator $\hat M$ is suppressed by $1/x^+$ but measurement of the radiation with finite $x^+$ provides us with complementary information of the shell's quantum state that we would otherwise miss by measuring only at $\mathcal{I}^+_R$.
Eigenvectors of the radiation can be computed, since $\sigma^-_{atm}$ is a self adjoint operator. If we restrict to the case where the shell's radiation is measured after the shell has collapsed $v<\sigma_m^+$, the measurements of the radiation projects us into the subspace of the Hilbert space given by the eigenvectors 
\begin{equation} \label{RadiationEigenstates}
R_k^{\sigma_m^+}(v)=\sqrt{\frac{2e^{\lambda v}}{\pi\hbar|1-e^{-\lambda(\sigma_m^+-v)}|}}\cos\Big(\frac{-p}{\lambda\hbar}e^{\lambda\sigma_m^+}\ln\Big|{1-e^{-\lambda (\sigma^+_m-v)}}\Big|\Big),
\end{equation} defined and normalized, as we said, for the region $v<\sigma_m^+$.
\begin{figure}[htp] \centering{
\includegraphics[scale=0.9]{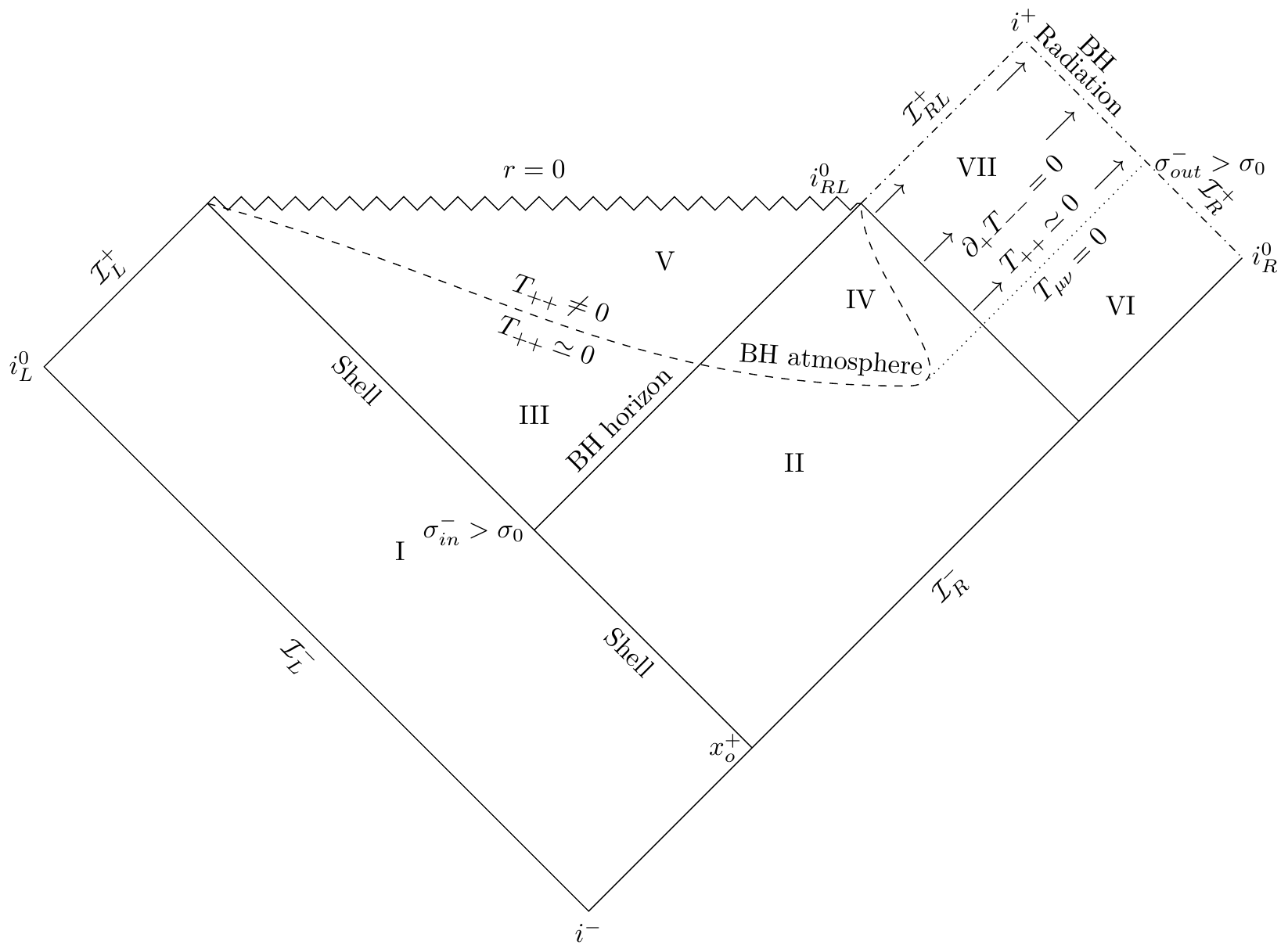}}
{\caption[r]{The description Fig. \ref{fig:PwoBR} fits in here, yet, if backreaction induces complete evaporation in finite time we would get a reduction of the atmosphere until it disappears.  
Region VI is flat, but VII is affected gravitationally by the Black Hole radiation.  
In fact, the exact calculation may also show the formation of a new black hole as the radiation reaches ${\cal I}^+_R$, an iterative process that we skip here. The apparition of spatial infinity $i^0_{RL}$ and $\mathcal{I}^+_{RL}$ after complete evaporation may be affected by the last bit of radiation of the black hole. Yet, we rather postpone the discussion until we get an analytical picture. All ambiguities in the description are represented by the dashdotted line.}}
\label{fig:PwBR}\end{figure} 
\subsection{Backreaction effects in Measurement \& Information}
 
We have seen that measurement of the density matrix at finite distance and time can lead to the determination of the shell's wavefunction. Yet it cannot be recovered using information exclusively on $\mathcal{I}^+_R$ without considering backreaction. 
We hypothesize  that when considering backreaction, the eigenstate of the Hawking radiation would not be constant on  $\mathcal{I}^+_R$ as is presently the case. This dependence would clearly come about in the case of complete evaporation of the black hole at certain finite $x^+_{evaporation}$. 
This is because the radiation would come exclusively from high curvature regions $x^+<x^+_{evaporation}$, which would cease to exist after evaporation. 
Let us consider briefly the atmosphere. 
In our approximation (\ref{atmosphere}) this region is clearly delimited by a Heaviside function, which is a substitute to the exact expression given for the Energy density flux from the conformal anomaly  at $\hbar $ order (\ref{WithoutAtmosphere}). 
For a null shell and no backreaction, from (\ref{ParticleFlux}), we conclude that the atmosphere would produce radiation only in the region where 
\begin{equation}\label{DashedLine}x_{in}^-(atm)\geq\frac{4Gp}{\lambda^2}(1- \frac{x^+_0}{x^+}).\end{equation} 
In this expression we can notice that the atmosphere grows until the radiation reaches ${\cal I}^+_R$ as in  Fig. \ref{fig:PwoBR}. Backreaction would change this behaviour at late times as in Fig. 2. The dashed line features the expected reduction of the atmosphere --by the introduction of corrections to $x^-_{in}(atm)$ in (\ref{DashedLine})-- and eventually meets the singularity at the point of complete evaporation. We do not provide conclusive evidence for the existence of the regions VI and VII in FIG. 2. This could be settled if it is proven that $x^+_{evaporation}<\infty$, which would imply to find the intersection between the singularity and the atmosphere's boundary. Yet, the complete evaporation is not a requisite for the information to reach $\mathcal{I}^+_R$.

We conclude arguing that the reduction of the atmosphere has important implications for the retrieval of information. In this scenario the eigenvectors of the outgoing radiation are of the form (\ref{RadiationEigenstates}) with $x^+_m=x^+(atm)$. This radiation reaches $\mathcal{I}^+_R$ unaltered, allowing us to retrieve different pieces of information about the shell at different times. This wouldn't be possible without backreaction because, as we saw in section \ref{measurements}, the eigenvectors of the radiation would be independent of the parameters in $\mathcal {I}^+_R$. This analysis about the effect of backreaction on the radiated information  is only valid in the two dimensional case. However, it suggests that it would not be possible to give a satisfactory answer to the information paradox until backreaction effects will be taken into account.

\section*{Aknowledgment}
The authors are very grateful towards Miguel Campiglia for countless and fruitful discussions. We would also like to thank Jorge Pullin for valuable inputs.

\bibliography{bibliography.bib}

\begin{thebibliography}{33}%
\makeatletter
\providecommand \@ifxundefined [1]{%
 \@ifx{#1\undefined}
}%
\providecommand \@ifnum [1]{%
 \ifnum #1\expandafter \@firstoftwo
 \else \expandafter \@secondoftwo
 \fi
}%
\providecommand \@ifx [1]{%
 \ifx #1\expandafter \@firstoftwo
 \else \expandafter \@secondoftwo
 \fi
}%
\providecommand \natexlab [1]{#1}%
\providecommand \enquote  [1]{``#1''}%
\providecommand \bibnamefont  [1]{#1}%
\providecommand \bibfnamefont [1]{#1}%
\providecommand \citenamefont [1]{#1}%
\providecommand \href@noop [0]{\@secondoftwo}%
\providecommand \href [0]{\begingroup \@sanitize@url \@href}%
\providecommand \@href[1]{\@@startlink{#1}\@@href}%
\providecommand \@@href[1]{\endgroup#1\@@endlink}%
\providecommand \@sanitize@url [0]{\catcode `\\12\catcode `\$12\catcode
  `\&12\catcode `\#12\catcode `\^12\catcode `\_12\catcode `\%12\relax}%
\providecommand \@@startlink[1]{}%
\providecommand \@@endlink[0]{}%
\providecommand \url  [0]{\begingroup\@sanitize@url \@url }%
\providecommand \@url [1]{\endgroup\@href {#1}{\urlprefix }}%
\providecommand \urlprefix  [0]{URL }%
\providecommand \Eprint [0]{\href }%
\providecommand \doibase [0]{http://dx.doi.org/}%
\providecommand \selectlanguage [0]{\@gobble}%
\providecommand \bibinfo  [0]{\@secondoftwo}%
\providecommand \bibfield  [0]{\@secondoftwo}%
\providecommand \translation [1]{[#1]}%
\providecommand \BibitemOpen [0]{}%
\providecommand \bibitemStop [0]{}%
\providecommand \bibitemNoStop [0]{.\EOS\space}%
\providecommand \EOS [0]{\spacefactor3000\relax}%
\providecommand \BibitemShut  [1]{\csname bibitem#1\endcsname}%
\let\auto@bib@innerbib\@empty
\bibitem [{\citenamefont {Hawking}(1975)}]{Hawking:1974sw}%
  \BibitemOpen
  \bibfield  {author} {\bibinfo {author} {\bibfnamefont {S.~W.}\ \bibnamefont
  {Hawking}},\ }\bibfield  {booktitle} {\emph {\bibinfo {booktitle} {{Euclidean
  quantum gravity}}},\ }\href {\doibase 10.1007/BF02345020, 10.1007/BF01608497}
  {\bibfield  {journal} {\bibinfo  {journal} {Commun. Math. Phys.}\ }\textbf
  {\bibinfo {volume} {43}},\ \bibinfo {pages} {199} (\bibinfo {year}
  {1975})}\BibitemShut {NoStop}%
\bibitem [{\citenamefont {Hajicek}(2003)}]{Hajicek:2002ny}%
  \BibitemOpen
  \bibfield  {author} {\bibinfo {author} {\bibfnamefont {P.}~\bibnamefont
  {Hajicek}},\ }\bibfield  {booktitle} {\emph {\bibinfo {booktitle}
  {{Proceedings, 271st WE-Heraeus Seminar on Aspects of Quantum Gravity: From
  Theory to Experiment Search: Bad Honnef, Germany, February 25-March 1,
  2002}}},\ }\href {\doibase 10.1007/978-3-540-45230-0_6} {\bibfield  {journal}
  {\bibinfo  {journal} {Lect. Notes Phys.}\ }\textbf {\bibinfo {volume}
  {631}},\ \bibinfo {pages} {255} (\bibinfo {year} {2003})},\ \Eprint
  {http://arxiv.org/abs/gr-qc/0204049} {arXiv:gr-qc/0204049 [gr-qc]}
  \BibitemShut {NoStop}%
\bibitem [{\citenamefont {Townsend}(1997)}]{Townsend:1997ku}%
  \BibitemOpen
  \bibfield  {author} {\bibinfo {author} {\bibfnamefont {P.~K.}\ \bibnamefont
  {Townsend}},\ }\href@noop {} {\  (\bibinfo {year} {1997})},\ \Eprint
  {http://arxiv.org/abs/gr-qc/9707012} {arXiv:gr-qc/9707012 [gr-qc]}
  \BibitemShut {NoStop}%
\bibitem [{\citenamefont {Almheiri}\ \emph {et~al.}(2013)\citenamefont
  {Almheiri}, \citenamefont {Marolf}, \citenamefont {Polchinski}, \citenamefont
  {Stanford},\ and\ \citenamefont {Sully}}]{Almheiri:2013hfa}%
  \BibitemOpen
  \bibfield  {author} {\bibinfo {author} {\bibfnamefont {A.}~\bibnamefont
  {Almheiri}}, \bibinfo {author} {\bibfnamefont {D.}~\bibnamefont {Marolf}},
  \bibinfo {author} {\bibfnamefont {J.}~\bibnamefont {Polchinski}}, \bibinfo
  {author} {\bibfnamefont {D.}~\bibnamefont {Stanford}}, \ and\ \bibinfo
  {author} {\bibfnamefont {J.}~\bibnamefont {Sully}},\ }\href {\doibase
  10.1007/JHEP09(2013)018} {\bibfield  {journal} {\bibinfo  {journal} {JHEP}\
  }\textbf {\bibinfo {volume} {09}},\ \bibinfo {pages} {018} (\bibinfo {year}
  {2013})},\ \Eprint {http://arxiv.org/abs/1304.6483} {arXiv:1304.6483
  [hep-th]} \BibitemShut {NoStop}%
\bibitem [{\citenamefont {Marolf}(2017)}]{Marolf:2017jkr}%
  \BibitemOpen
  \bibfield  {author} {\bibinfo {author} {\bibfnamefont {D.}~\bibnamefont
  {Marolf}},\ }\href {\doibase 10.1088/1361-6633/aa77cc} {\bibfield  {journal}
  {\bibinfo  {journal} {Rept. Prog. Phys.}\ }\textbf {\bibinfo {volume} {80}},\
  \bibinfo {pages} {092001} (\bibinfo {year} {2017})},\ \Eprint
  {http://arxiv.org/abs/1703.02143} {arXiv:1703.02143 [gr-qc]} \BibitemShut
  {NoStop}%
\bibitem [{\citenamefont {Unruh}\ and\ \citenamefont
  {Wald}(2017)}]{Unruh:2017uaw}%
  \BibitemOpen
  \bibfield  {author} {\bibinfo {author} {\bibfnamefont {W.~G.}\ \bibnamefont
  {Unruh}}\ and\ \bibinfo {author} {\bibfnamefont {R.~M.}\ \bibnamefont
  {Wald}},\ }\href {\doibase 10.1088/1361-6633/aa778e} {\bibfield  {journal}
  {\bibinfo  {journal} {Rept. Prog. Phys.}\ }\textbf {\bibinfo {volume} {80}},\
  \bibinfo {pages} {092002} (\bibinfo {year} {2017})},\ \Eprint
  {http://arxiv.org/abs/1703.02140} {arXiv:1703.02140 [hep-th]} \BibitemShut
  {NoStop}%
\bibitem [{\citenamefont {Chakraborty}\ and\ \citenamefont
  {Lochan}(2017)}]{Chakraborty:2017pmn}%
  \BibitemOpen
  \bibfield  {author} {\bibinfo {author} {\bibfnamefont {S.}~\bibnamefont
  {Chakraborty}}\ and\ \bibinfo {author} {\bibfnamefont {K.}~\bibnamefont
  {Lochan}},\ }\href {\doibase 10.3390/universe3030055} {\bibfield  {journal}
  {\bibinfo  {journal} {Universe}\ }\textbf {\bibinfo {volume} {3}},\ \bibinfo
  {pages} {55} (\bibinfo {year} {2017})},\ \Eprint
  {http://arxiv.org/abs/1702.07487} {arXiv:1702.07487 [gr-qc]} \BibitemShut
  {NoStop}%
\bibitem [{\citenamefont {Kuchar}(1994)}]{Kuchar:1994zk}%
  \BibitemOpen
  \bibfield  {author} {\bibinfo {author} {\bibfnamefont {K.~V.}\ \bibnamefont
  {Kuchar}},\ }\href {\doibase 10.1103/PhysRevD.50.3961} {\bibfield  {journal}
  {\bibinfo  {journal} {Phys. Rev.}\ }\textbf {\bibinfo {volume} {D50}},\
  \bibinfo {pages} {3961} (\bibinfo {year} {1994})},\ \Eprint
  {http://arxiv.org/abs/gr-qc/9403003} {arXiv:gr-qc/9403003 [gr-qc]}
  \BibitemShut {NoStop}%
\bibitem [{\citenamefont {Eyheralde}\ \emph {et~al.}(2017)\citenamefont
  {Eyheralde}, \citenamefont {Campiglia}, \citenamefont {Gambini},\ and\
  \citenamefont {Pullin}}]{Eyheralde:2017jzd}%
  \BibitemOpen
  \bibfield  {author} {\bibinfo {author} {\bibfnamefont {R.}~\bibnamefont
  {Eyheralde}}, \bibinfo {author} {\bibfnamefont {M.}~\bibnamefont
  {Campiglia}}, \bibinfo {author} {\bibfnamefont {R.}~\bibnamefont {Gambini}},
  \ and\ \bibinfo {author} {\bibfnamefont {J.}~\bibnamefont {Pullin}},\ }\href
  {\doibase 10.1088/1361-6382/aa8e30} {\bibfield  {journal} {\bibinfo
  {journal} {Class. Quant. Grav.}\ }\textbf {\bibinfo {volume} {34}},\ \bibinfo
  {pages} {235015} (\bibinfo {year} {2017})},\ \Eprint
  {http://arxiv.org/abs/1705.05722} {arXiv:1705.05722 [gr-qc]} \BibitemShut
  {NoStop}%
\bibitem [{\citenamefont {Callan}\ \emph {et~al.}(1992)\citenamefont {Callan},
  \citenamefont {Giddings}, \citenamefont {Harvey},\ and\ \citenamefont
  {Strominger}}]{Callan:1992rs}%
  \BibitemOpen
  \bibfield  {author} {\bibinfo {author} {\bibfnamefont {C.~G.}\ \bibnamefont
  {Callan}, \bibfnamefont {Jr.}}, \bibinfo {author} {\bibfnamefont {S.~B.}\
  \bibnamefont {Giddings}}, \bibinfo {author} {\bibfnamefont {J.~A.}\
  \bibnamefont {Harvey}}, \ and\ \bibinfo {author} {\bibfnamefont
  {A.}~\bibnamefont {Strominger}},\ }\href {\doibase 10.1103/PhysRevD.45.R1005}
  {\bibfield  {journal} {\bibinfo  {journal} {Phys. Rev.}\ }\textbf {\bibinfo
  {volume} {D45}},\ \bibinfo {pages} {R1005} (\bibinfo {year} {1992})},\
  \Eprint {http://arxiv.org/abs/hep-th/9111056} {arXiv:hep-th/9111056}
  \BibitemShut {NoStop}%
\bibitem [{\citenamefont {Giddings}\ and\ \citenamefont
  {Nelson}(1992)}]{Giddings:1992ff}%
  \BibitemOpen
  \bibfield  {author} {\bibinfo {author} {\bibfnamefont {S.~B.}\ \bibnamefont
  {Giddings}}\ and\ \bibinfo {author} {\bibfnamefont {W.~M.}\ \bibnamefont
  {Nelson}},\ }\href {\doibase 10.1103/PhysRevD.46.2486} {\bibfield  {journal}
  {\bibinfo  {journal} {Phys. Rev.}\ }\textbf {\bibinfo {volume} {D46}},\
  \bibinfo {pages} {2486} (\bibinfo {year} {1992})},\ \Eprint
  {http://arxiv.org/abs/hep-th/9204072} {arXiv:hep-th/9204072 [hep-th]}
  \BibitemShut {NoStop}%
\bibitem [{\citenamefont {Strominger}(1994)}]{Strominger:1994tn}%
  \BibitemOpen
  \bibfield  {author} {\bibinfo {author} {\bibfnamefont {A.}~\bibnamefont
  {Strominger}},\ }in\ \href@noop {} {\emph {\bibinfo {booktitle} {{NATO
  Advanced Study Institute: Les Houches Summer School, Session 62: Fluctuating
  Geometries in Statistical Mechanics and Field Theory Les Houches, France,
  August 2-September 9, 1994}}}}\ (\bibinfo {year} {1994})\ \Eprint
  {http://arxiv.org/abs/hep-th/9501071} {hep-th/9501071} \BibitemShut {NoStop}%
\bibitem [{\citenamefont {Schoutens}\ \emph {et~al.}(1993)\citenamefont
  {Schoutens}, \citenamefont {Verlinde},\ and\ \citenamefont
  {Verlinde}}]{Schoutens:1993hu}%
  \BibitemOpen
  \bibfield  {author} {\bibinfo {author} {\bibfnamefont {K.}~\bibnamefont
  {Schoutens}}, \bibinfo {author} {\bibfnamefont {H.~L.}\ \bibnamefont
  {Verlinde}}, \ and\ \bibinfo {author} {\bibfnamefont {E.~P.}\ \bibnamefont
  {Verlinde}},\ }\href {\doibase 10.1103/PhysRevD.48.2670} {\bibfield
  {journal} {\bibinfo  {journal} {Phys. Rev.}\ }\textbf {\bibinfo {volume}
  {D48}},\ \bibinfo {pages} {2670} (\bibinfo {year} {1993})},\ \Eprint
  {http://arxiv.org/abs/hep-th/9304128} {arXiv:hep-th/9304128 [hep-th]}
  \BibitemShut {NoStop}%
\bibitem [{\citenamefont {Varadarajan}(1995)}]{Varadarajan:1995jj}%
  \BibitemOpen
  \bibfield  {author} {\bibinfo {author} {\bibfnamefont {M.}~\bibnamefont
  {Varadarajan}},\ }\href {\doibase 10.1103/PhysRevD.52.7080} {\bibfield
  {journal} {\bibinfo  {journal} {Phys. Rev.}\ }\textbf {\bibinfo {volume}
  {D52}},\ \bibinfo {pages} {7080} (\bibinfo {year} {1995})},\ \Eprint
  {http://arxiv.org/abs/gr-qc/9508039} {arXiv:gr-qc/9508039 [gr-qc]}
  \BibitemShut {NoStop}%
\bibitem [{\citenamefont {Ashtekar}\ \emph {et~al.}(2008)\citenamefont
  {Ashtekar}, \citenamefont {Taveras},\ and\ \citenamefont
  {Varadarajan}}]{Ashtekar:2008jd}%
  \BibitemOpen
  \bibfield  {author} {\bibinfo {author} {\bibfnamefont {A.}~\bibnamefont
  {Ashtekar}}, \bibinfo {author} {\bibfnamefont {V.}~\bibnamefont {Taveras}}, \
  and\ \bibinfo {author} {\bibfnamefont {M.}~\bibnamefont {Varadarajan}},\
  }\href {\doibase 10.1103/PhysRevLett.100.211302} {\bibfield  {journal}
  {\bibinfo  {journal} {Phys. Rev. Lett.}\ }\textbf {\bibinfo {volume} {100}},\
  \bibinfo {pages} {211302} (\bibinfo {year} {2008})},\ \Eprint
  {http://arxiv.org/abs/0801.1811} {arXiv:0801.1811 [gr-qc]} \BibitemShut
  {NoStop}%
\bibitem [{\citenamefont {Ashtekar}\ \emph {et~al.}(2011)\citenamefont
  {Ashtekar}, \citenamefont {Pretorius},\ and\ \citenamefont
  {Ramazanoglu}}]{Ashtekar:2010qz}%
  \BibitemOpen
  \bibfield  {author} {\bibinfo {author} {\bibfnamefont {A.}~\bibnamefont
  {Ashtekar}}, \bibinfo {author} {\bibfnamefont {F.}~\bibnamefont {Pretorius}},
  \ and\ \bibinfo {author} {\bibfnamefont {F.~M.}\ \bibnamefont
  {Ramazanoglu}},\ }\href {\doibase 10.1103/PhysRevD.83.044040} {\bibfield
  {journal} {\bibinfo  {journal} {Phys. Rev.}\ }\textbf {\bibinfo {volume}
  {D83}},\ \bibinfo {pages} {044040} (\bibinfo {year} {2011})},\ \Eprint
  {http://arxiv.org/abs/1012.0077} {arXiv:1012.0077 [gr-qc]} \BibitemShut
  {NoStop}%
\bibitem [{\citenamefont {Modak}\ and\ \citenamefont
  {Sudarsky}(2017)}]{Modak:2016uwr}%
  \BibitemOpen
  \bibfield  {author} {\bibinfo {author} {\bibfnamefont {S.~K.}\ \bibnamefont
  {Modak}}\ and\ \bibinfo {author} {\bibfnamefont {D.}~\bibnamefont
  {Sudarsky}},\ }\href {\doibase 10.1007/978-3-319-51700-1_18} {\bibfield
  {journal} {\bibinfo  {journal} {Fundam. Theor. Phys.}\ }\textbf {\bibinfo
  {volume} {187}},\ \bibinfo {pages} {303} (\bibinfo {year} {2017})},\ \Eprint
  {http://arxiv.org/abs/1607.05410} {arXiv:1607.05410 [gr-qc]} \BibitemShut
  {NoStop}%
\bibitem [{\citenamefont {Guo}\ and\ \citenamefont {Cai}(2018)}]{Guo:2018cgy}%
  \BibitemOpen
  \bibfield  {author} {\bibinfo {author} {\bibfnamefont {X.-K.}\ \bibnamefont
  {Guo}}\ and\ \bibinfo {author} {\bibfnamefont {Q.-Y.}\ \bibnamefont {Cai}},\
  }\href {\doibase 10.1142/S0217732318501031} {\bibfield  {journal} {\bibinfo
  {journal} {Modern Physics Letters A}\ }\textbf {\bibinfo {volume} {33}},\
  \bibinfo {pages} {1850103} (\bibinfo {year} {2018})},\ \Eprint
  {http://arxiv.org/abs/https://doi.org/10.1142/S0217732318501031}
  {https://doi.org/10.1142/S0217732318501031} \BibitemShut {NoStop}%
\bibitem [{\citenamefont {Ho}\ and\ \citenamefont {Matsuo}(2018)}]{Ho:2018jkm}%
  \BibitemOpen
  \bibfield  {author} {\bibinfo {author} {\bibfnamefont {P.-M.}\ \bibnamefont
  {Ho}}\ and\ \bibinfo {author} {\bibfnamefont {Y.}~\bibnamefont {Matsuo}},\
  }\href@noop {} {\  (\bibinfo {year} {2018})},\ \Eprint
  {http://arxiv.org/abs/1804.04821} {arXiv:1804.04821 [hep-th]} \BibitemShut
  {NoStop}%
\bibitem [{\citenamefont {Louko}\ \emph {et~al.}(1998)\citenamefont {Louko},
  \citenamefont {Whiting},\ and\ \citenamefont {Friedman}}]{Louko:1997wc}%
  \BibitemOpen
  \bibfield  {author} {\bibinfo {author} {\bibfnamefont {J.}~\bibnamefont
  {Louko}}, \bibinfo {author} {\bibfnamefont {B.~F.}\ \bibnamefont {Whiting}},
  \ and\ \bibinfo {author} {\bibfnamefont {J.~L.}\ \bibnamefont {Friedman}},\
  }\href {\doibase 10.1103/PhysRevD.57.2279} {\bibfield  {journal} {\bibinfo
  {journal} {Phys. Rev.}\ }\textbf {\bibinfo {volume} {D57}},\ \bibinfo {pages}
  {2279} (\bibinfo {year} {1998})},\ \Eprint
  {http://arxiv.org/abs/gr-qc/9708012} {arXiv:gr-qc/9708012 [gr-qc]}
  \BibitemShut {NoStop}%
\bibitem [{\citenamefont {Campiglia}\ \emph {et~al.}(2016)\citenamefont
  {Campiglia}, \citenamefont {Gambini}, \citenamefont {Olmedo},\ and\
  \citenamefont {Pullin}}]{Campiglia:2016fzp}%
  \BibitemOpen
  \bibfield  {author} {\bibinfo {author} {\bibfnamefont {M.}~\bibnamefont
  {Campiglia}}, \bibinfo {author} {\bibfnamefont {R.}~\bibnamefont {Gambini}},
  \bibinfo {author} {\bibfnamefont {J.}~\bibnamefont {Olmedo}}, \ and\ \bibinfo
  {author} {\bibfnamefont {J.}~\bibnamefont {Pullin}},\ }\href {\doibase
  10.1088/0264-9381/33/18/18LT01} {\bibfield  {journal} {\bibinfo  {journal}
  {Class. Quant. Grav.}\ }\textbf {\bibinfo {volume} {33}},\ \bibinfo {pages}
  {18LT01} (\bibinfo {year} {2016})},\ \Eprint
  {http://arxiv.org/abs/1601.05688} {arXiv:1601.05688 [gr-qc]} \BibitemShut
  {NoStop}%
\bibitem [{\citenamefont {Rastgoo}(2013)}]{Rastgoo:2013isa}%
  \BibitemOpen
  \bibfield  {author} {\bibinfo {author} {\bibfnamefont {S.}~\bibnamefont
  {Rastgoo}},\ }\href@noop {} {\  (\bibinfo {year} {2013})},\ \Eprint
  {http://arxiv.org/abs/1304.7836} {arXiv:1304.7836 [gr-qc]} \BibitemShut
  {NoStop}%
\bibitem [{\citenamefont {Corichi}\ \emph {et~al.}(2016)\citenamefont
  {Corichi}, \citenamefont {Olmedo},\ and\ \citenamefont
  {Rastgoo}}]{Corichi:2016nkp}%
  \BibitemOpen
  \bibfield  {author} {\bibinfo {author} {\bibfnamefont {A.}~\bibnamefont
  {Corichi}}, \bibinfo {author} {\bibfnamefont {J.}~\bibnamefont {Olmedo}}, \
  and\ \bibinfo {author} {\bibfnamefont {S.}~\bibnamefont {Rastgoo}},\ }\href
  {\doibase 10.1103/PhysRevD.94.084050} {\bibfield  {journal} {\bibinfo
  {journal} {Phys. Rev.}\ }\textbf {\bibinfo {volume} {D94}},\ \bibinfo {pages}
  {084050} (\bibinfo {year} {2016})},\ \Eprint
  {http://arxiv.org/abs/1608.06246} {arXiv:1608.06246 [gr-qc]} \BibitemShut
  {NoStop}%
\bibitem [{\citenamefont {Gambini}\ and\ \citenamefont
  {Pullin}(2008)}]{Gambini:2008dy}%
  \BibitemOpen
  \bibfield  {author} {\bibinfo {author} {\bibfnamefont {R.}~\bibnamefont
  {Gambini}}\ and\ \bibinfo {author} {\bibfnamefont {J.}~\bibnamefont
  {Pullin}},\ }\href {\doibase 10.1103/PhysRevLett.101.161301} {\bibfield
  {journal} {\bibinfo  {journal} {Phys. Rev. Lett.}\ }\textbf {\bibinfo
  {volume} {101}},\ \bibinfo {pages} {161301} (\bibinfo {year} {2008})},\
  \Eprint {http://arxiv.org/abs/0805.1187} {arXiv:0805.1187 [gr-qc]}
  \BibitemShut {NoStop}%
\bibitem [{\citenamefont {Weisstein}()}]{GammaLowfreq:Online}%
  \BibitemOpen
  \bibfield  {author} {\bibinfo {author} {\bibfnamefont {E.~W.}\ \bibnamefont
  {Weisstein}},\ }\href@noop {} {\enquote {\bibinfo {title}
  {\href{http://mathworld.wolfram.com/GammaFunction.html}{Gamma Function}},}\
  }\bibinfo {howpublished} {From MathWorld--A Wolfram Web Resource}\BibitemShut
  {NoStop}%
\bibitem [{\citenamefont
  {\href{https://math.stackexchange.com/users/21783/raymond-manzoni}{Raymond
  Manzoni}}()}]{GammaHighFreq:Online}%
  \BibitemOpen
  \bibfield  {author} {\bibinfo {author} {\bibnamefont
  {\href{https://math.stackexchange.com/users/21783/raymond-manzoni}{Raymond
  Manzoni}}},\ }\href@noop {} {\enquote {\bibinfo {title}
  {\href{https://math.stackexchange.com/q/456010}{Real and imaginary part of
  Gamma function}},}\ }\bibinfo {howpublished} {Mathematics Stack Exchange},\
  \Eprint {http://arxiv.org/abs/https://math.stackexchange.com/q/456010}
  {https://math.stackexchange.com/q/456010} \BibitemShut {NoStop}%
\bibitem [{\citenamefont {Mecklenbr{\"a}uker}\ and\ \citenamefont
  {Hlawatsch}(1997)}]{mecklenbräuker1997wigner}%
  \BibitemOpen
  \bibfield  {author} {\bibinfo {author} {\bibfnamefont {W.}~\bibnamefont
  {Mecklenbr{\"a}uker}}\ and\ \bibinfo {author} {\bibfnamefont
  {F.}~\bibnamefont {Hlawatsch}},\ }\href
  {https://books.google.com.uy/books?id=NPZSAAAAMAAJ} {\emph {\bibinfo {title}
  {The Wigner Distribution: Theory and Applications in Signal Processing}}}\
  (\bibinfo  {publisher} {Elsevier Science},\ \bibinfo {year}
  {1997})\BibitemShut {NoStop}%
\bibitem [{\citenamefont {Wong}(2001)}]{wong2001asymptotic}%
  \BibitemOpen
  \bibfield  {author} {\bibinfo {author} {\bibfnamefont {R.}~\bibnamefont
  {Wong}},\ }\href {https://books.google.com.uy/books?id=KQHPHPZs8k4C} {\emph
  {\bibinfo {title} {Asymptotic Approximation of Integrals}}},\ Classics in
  Applied Mathematics\ (\bibinfo  {publisher} {Society for Industrial and
  Applied Mathematics},\ \bibinfo {year} {2001})\BibitemShut {NoStop}%
\bibitem [{\citenamefont {Cohen}(1995)}]{cohen1995time}%
  \BibitemOpen
  \bibfield  {author} {\bibinfo {author} {\bibfnamefont {L.}~\bibnamefont
  {Cohen}},\ }\href {https://books.google.com.uy/books?id=CSKLQgAACAAJ} {\emph
  {\bibinfo {title} {Time-frequency Analysis}}},\ Electrical engineering signal
  processing\ (\bibinfo  {publisher} {Prentice Hall PTR},\ \bibinfo {year}
  {1995})\BibitemShut {NoStop}%
\bibitem [{\citenamefont {Parker}\ and\ \citenamefont
  {Toms}(2009)}]{Parker:2009uva}%
  \BibitemOpen
  \bibfield  {author} {\bibinfo {author} {\bibfnamefont {L.~E.}\ \bibnamefont
  {Parker}}\ and\ \bibinfo {author} {\bibfnamefont {D.}~\bibnamefont {Toms}},\
  }\href {\doibase 10.1017/CBO9780511813924} {\emph {\bibinfo {title} {{Quantum
  Field Theory in Curved Spacetime}}}},\ Cambridge Monographs on Mathematical
  Physics\ (\bibinfo  {publisher} {Cambridge University Press},\ \bibinfo
  {year} {2009})\BibitemShut {NoStop}%
\bibitem [{\citenamefont {Schwartz}(2014)}]{Schwartz:2013pla}%
  \BibitemOpen
  \bibfield  {author} {\bibinfo {author} {\bibfnamefont {M.~D.}\ \bibnamefont
  {Schwartz}},\ }\href
  {http://www.cambridge.org/us/academic/subjects/physics/theoretical-physics-and-mathematical-physics/quantum-field-theory-and-standard-model}
  {\emph {\bibinfo {title} {{Quantum Field Theory and the Standard Model}}}}\
  (\bibinfo  {publisher} {Cambridge University Press},\ \bibinfo {year}
  {2014})\BibitemShut {NoStop}%
\bibitem [{\citenamefont {Birrell}\ and\ \citenamefont
  {Davies}(1984)}]{Birrell:1982ix}%
  \BibitemOpen
  \bibfield  {author} {\bibinfo {author} {\bibfnamefont {N.~D.}\ \bibnamefont
  {Birrell}}\ and\ \bibinfo {author} {\bibfnamefont {P.~C.~W.}\ \bibnamefont
  {Davies}},\ }\href {\doibase 10.1017/CBO9780511622632} {\emph {\bibinfo
  {title} {{Quantum Fields in Curved Space}}}},\ Cambridge Monographs on
  Mathematical Physics\ (\bibinfo  {publisher} {Cambridge Univ. Press},\
  \bibinfo {address} {Cambridge, UK},\ \bibinfo {year} {1984})\BibitemShut
  {NoStop}%
\bibitem [{\citenamefont {Fabbri}\ and\ \citenamefont
  {Navarro-Salas}(2005)}]{Fabbri:2005mw}%
  \BibitemOpen
  \bibfield  {author} {\bibinfo {author} {\bibfnamefont {A.}~\bibnamefont
  {Fabbri}}\ and\ \bibinfo {author} {\bibfnamefont {J.}~\bibnamefont
  {Navarro-Salas}},\ }\href {\doibase 10.1142/9781860947223_0001} {\emph
  {\bibinfo {title} {{Modeling black hole evaporation}}}}\ (\bibinfo
  {publisher} {Imperial College Press},\ \bibinfo {year} {2005})\BibitemShut
  {NoStop}%
\end{thebibliography}%



\end{document}